# Quantum Nanophotonics with Group IV defects in Diamond

*Carlo Bradac,*[1,*] *Weibo Gao,*[2] *Jacopo Forneris,*[3] *Matt Trusheim*[4] & *Igor Aharonovich*[1]

[1] School of Mathematical and Physical Sciences, Faculty of Science, University of Technology Sydney, Australia 2007
[2] Division of Physics and Applied Physics, School of Physical and Mathematical Sciences, Nanyang Technological University, Singapore 637371, Singapore
[3] Istituto Nazionale di Fisica Nucleare (INFN), Torino, 10125, Italy
[4] Department of Electrical Engineering and Computer Science, Massachusetts Institute of Technology, Cambridge, Massachusetts 02139, USA

* Corresponding author, Dr Carlo Bradac: carlo.bradac@uts.edu.au

**Abstract**
Diamond photonics is an ever-growing field of research driven by the prospects of harnessing diamond and its colour centres as suitable hardware for solid-state quantum applications. The last two decades have seen the field been shaped by the nitrogen-vacancy (NV) centre both with breakthrough fundamental physics demonstrations and practical realizations. Recently however, an entire suite of other diamond defects has emerged. They are *M-V* colour centres, where *M* indicates one of the elements in the IV column of the periodic table—Si, Ge, Sn and Pb, and *V* indicates lattice vacancies, i.e. missing next-neighbour carbon atoms. These centres exhibit a much stronger emission into the zero-phonon line then the NV centre, they display inversion symmetry, and can be engineered using ion implantation—all attractive features for scalable quantum photonic architectures based on solid-state, single-photon sources. In this perspective, we highlight the leading techniques for engineering and characterizing these diamond defects, discuss the current state-of-the-art of group IV-based devices and provide an outlook of the future directions the field is taking towards the realisation of solid-state quantum photonics with diamond.

**Introduction**
The field of diamond photonics is marching into its third decade—its birth arguably marked by the 1997 discovery of room-temperature optically detected magnetic resonance (ODMR) from a single diamond nitrogen-vacancy (NV) centre.[1] The unique ability of the NV's spin to be initialised, manipulated and optically read out at room temperature gave substance to the aspiration of realising solid-state quantum bits operating in ambient conditions.[2,3] Tremendous efforts followed, driven by the goal to engineer high-quality NV centres with long spin coherence times, and ameliorate the fabrication of diamond nanostructures for efficient light extraction.[4-7] The remarkable progress made in pursuit of this endeavour resulted in landmark realisations both in fundamental and applied science including on-demand entanglement,[3] nanoscale nuclear magnetic resonance[8,9] and quantum memories.[10]

However for applications requiring better photon throughput, e.g. quantum repeaters, the NV centre is not ideal. Its long fluorescent lifetime (~11 ns) and weak emission into the zero-phonon line (only ~4% at room temperature), puts an upper bound to the maximum photon rates achievable when employing NV centres in basic quantum photonic devices. While significant progress has been made towards improving photon extraction—for instance via the use of solid immersion lenses, diamond antennas or optical resonators—having photon emitters with better-suited properties is a desirable alternative.

In the recent past, other colour centres in diamond have thus been explored. Initial experiments revealed a diverse spread of narrow-band emitters, spanning over the visible and near infrared spectral range.[11] The silicon-vacancy (SiV⁻) colour centre was the first out of the group IV atoms to be investigated, as it was a known colour centre in diamond from the early 80s[12]—yet only unambiguously identified as a silicon related defect with its current electronic level structure in the 90s.[13, 14] Unfortunately, the first studies on single SiV⁻ defects created by ion implantation gave mixed results.[15] While the centre showed to possess a sharp zero-phonon line (ZPL) at 738 nm (responsible for >70% of the total emission) with only weak vibronic sidebands at room temperature, and a short photoluminescence lifetime of ~1 ns, it also displayed low single-photon emission rates (a few kcounts/s) and low radiative quantum yield (~0.05).

In 2011, the centre was revisited after a study from Neu et al.[16] showed ultra-bright emission (~4.8 Mcounts/s, at saturation) from single diamond SiV⁻ defects grown via microwave-plasma-assisted chemical vapour deposition on an iridium substrate. The discovery reinvigorated the interest for this defect, which resulted into better understanding of its level structure,[17, 18] photophysical properties[19, 20] and spin coherence times, as well as driving the design of schemes for initialization, readout, and coherent preparation of the centre.[21, 22] Remarkably, the SiV⁻ centre shows nearly lifetime-broadened optical emission, in both nanodiamonds[23, 24] and bulk crystals.[25] This derives from its inversion symmetry (group $D_{3d}$, with the silicon atom in a split-vacancy configuration) protecting the optical transitions from local electric field fluctuations,[26] which in turn allows for the existence—and potentially fabrication—of multiple, intrinsically-identical emitters in high-quality bulk diamond.[27] However, one of the main drawbacks of the SiV⁻ is its aforementioned, intrinsically low quantum efficiency—alongside sub-microsecond spin coherence time even at cryogenic temperatures.[21, 28] This prompted researchers around the world to explore—naturally—other group IV elements foreseeing that they will form colour centres with the same symmetry, and anticipating some of these might have a combination of desired properties with no or limited shortcomings. In 2015, the germanium-vacancy (GeV) centre was identified,[29-31] followed by the tin-vacancy (SnV)[32, 33] and the led-vacancy (PbV) centres.[34, 35] The main properties of these defects are summarised in Table 1 and their structure and level diagram are shown in Figure 1.

| | ZPL | FHWM at RT | Excited state Lifetime | Ground state splitting | Spin–lattice relaxation $T_1$ | Transverse relaxation time T2 |
|---|---|---|---|---|---|---|
| SiV⁻ | 738 nm | 0.7-5 nm | 1.0 - 2.4 ns | ~ 50 GHz | > 1s at 100 mK | $T_2$ ~13 ms at 100 mK |
| GeV⁻ | 602 nm | ~ 5 nm | 1.4–5.5 ns | ~ 170 GHz | 25 μs at 2 K | $T_2^*$ ~ 20 ns |
| SnV⁻ | 620 nm | ~ 6 nm | ~ 6 ns | ~ 850 GHz | ~60 μs | N/A |
| PbV⁻ | 520 nm, 552 nm | ~7 nm | > 3 ns | 5.7 THz @ 520 nm 4.2 THz @ 552 nm | N/A | N/A |

**Table 1.** Photophysical properties of the group IV defects

This perspective highlights the progress made along this journey—the challenges met and the successes made while engineering and characterising new group-IV emitters, as well as testing their performance for the realization of scalable quantum photonic devices. We first present the general properties of these emitters and the methods developed to engineer them. We then discuss the spin coherence properties of the SiV⁻ and the GeV⁻ centres, and show the current state of the art for a few representative quantum photonic experiments realised with these emitters. We conclude by giving a broad outlook over the outstanding obstacles, which might focus future directions and advancements in the field.

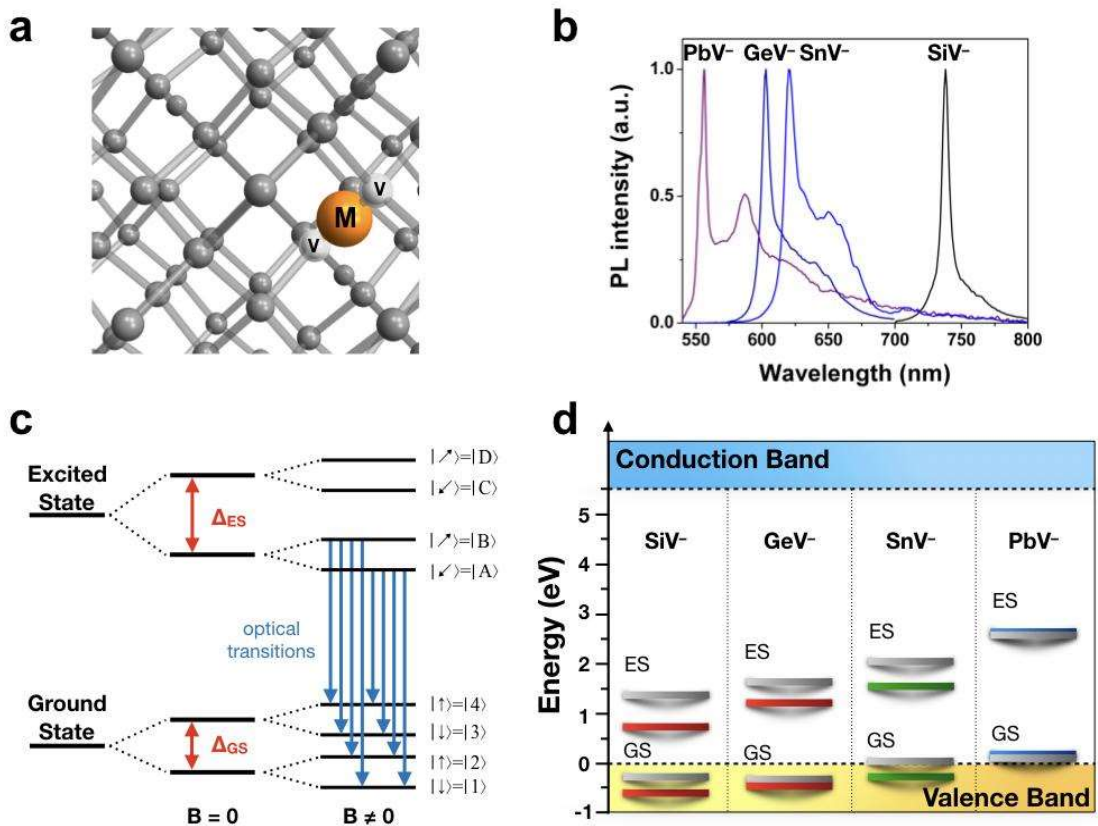

**Figure 1.** Photoluminescence of group-IV centres in diamond. **a)** Atomic structure of group-IV colour centres in the $D_{3d}$ symmetry, split-vacancy configuration. The group-IV element (M, orange) lies between two nearest-neighbours missing carbon atoms (V, white). **b)** Typical room-temperature photoluminescence (PL) spectrum of SiV⁻, GeV⁻, SnV⁻ and PbV⁻ centres. **c)** Energy structure of impurity-divacancy centres within zero and non-zero external magnetic field B. **d)** Predicted energies of the ground (GS) and excited (ES) states of the group-IV colour centres with respect to the diamond valence band maximum (i.e. the location of the valence band at the Brillouin-zone centre). Levels in grey, red, green, blue refer to values calculated in Refs.[29, 32, 34, 36]

**Defects fabrication strategies**

The photophysical properties of diamond M-V centres are uniquely favourable to both fundamental and practical realizations in quantum photonics. As a result, one major aspect, which is currently shaping the field, is the development of methods for the deterministic—ideally scalable—fabrication of these defects. Group-IV emitters in diamond are most commonly produced through either impurity incorporation during synthesis or via targeted ion implantation.

**Incorporation**. The formation of SiV⁻ centres in diamond is rather simple. Due to the residual silicon, often present in chemical vapor deposition (CVD) chambers from quartz bell jars or substrates, most of CVD-grown nanodiamond does contain SiV⁻ centres. Likewise, many high pressure high temperature (HPHT) nanodiamonds and bulk single crystals can include SiV⁻ centres that are present due to addition of silicon-containing precursors.[37] Similarly, the incorporation of GeV⁻ centres in diamond has been demonstrated by utilizing suitable precursors or substrates both during HPHT and CVD synthesis, while SnV⁻ has only been synthesized via HPHT synthesis, so far.[38-41] The effectiveness of the synthesis depends on the solubility of the specific element, and thus decreases as the atomic radius increases. It should be noted that the synthesized emitters generally exhibit better optical properties, e.g. higher photostability and narrower emission bandwidth, than those of centres fabricated by ion implantation—mainly owing to the higher crystalline quality of the former.[24, 42] However, this approach has a significant drawback in the context of device fabrication—the synthesis process only offers statistical control on the number of emitters and does not allow for precise positioning. For this reason, incorporation during synthesis is arguably best suited for hybrid approaches involving the fabrication of emitters in, for instance, individual diamond nanoparticles which are subsequently integrated with external photonic structures by pick-and-place techniques.[43]

**Ion implantation.** This approach stems from an enticing perspective: to individually control the number and position of the colour centres in the diamond material—ideally forming arbitrary arrays of quantum emitters. Ion implantation is a technologically-mature method which inherits a large body of knowledge from silicon-based electronic device manufacturing. Its exploitation for quantum device fabrication though is still limited for it currently lacks reliable, sub-50 nm positioning accuracy, single-ion delivery, and close-to-unity formation yield (i.e. ratio between optically-active emitters and number of implanted ions).[44]

*Spatial accuracy.* The deterministic formation of colour centres in diamond via implantation is intrinsically limited by lateral straggling, i.e. scattering of the ions as they travel through the target material. Notably, the relatively high mass of group-IV ions lessens the effects of straggling in diamond enabling sub-30 nm implantation accuracy with energies that, for instance for Si ions, can be as high as 200 keV. Higher placement accuracies (<10 nm) can be achieved at lower energies and for heavier elements (Ge–Pb).[45] Ion straggling is only one of the factors limiting spatial accuracy. The diameter of the impinging ion beam and the diffusion length of the atoms during the thermal annealing step employed to form the vacancy-centre complexes play a prime role too (see below).

Arguably the main outstanding challenge to deterministic nanoscale positioning of colour centres in diamond is thus the availability of specific beam optics for accurate spatial delivery of individual ions. To this end, two main approaches are being pursued—both currently achieving resolutions well-below 100 nm—nanoscale focusing and scanning of a rarefied ion beam via electromagnetic lenses, and use of collimators in combination with nano-apertures. For the first approach, the main obstacle is the limited focusing capability of conventional MeV energy accelerators. This has however been overcome by the development—still ongoing— of custom focused ion beam (FIB) apparatuses with commercially available alternatives to ordinary gallium liquid metal ion sources.[46-49] Conversely—relevant to the second approach— achromatic nanoscale collimators are compatible with any ion species as well as energy, and have been used in a variety of configurations to achieve spatially-resolved ion implantation. Specifically, collimators can consist of fixed apertures registered to specific implantation targets on the substrate,[50] or can be mounted on a scanning system enabling ion delivery at arbitrary positions.[51]

*Single ion detection.* In general, the delivery of a predefined number of ions on the diamond target cannot be achieved by a Poissonian approach which relies on preset exposure time and ion current. A much more deterministic solution is needed. Real-time counting of individual ions as they strike the target is one such solution. In this case, the target substrate itself is engineered to be a solid-state particle detector via patterning of suitable electrode geometries.[52] This strategy also allows, in perspective, the exploitation of the same electrodes for subsequent applications, such as inducing stark shift and electroluminescence. However, single-ion detection is challenged by the high energy required (~13 eV) to generate an electron-hole pair in diamond. To date, the detection of Si ions at energies as low as 200 keV has been demonstrated using single-ion counting techniques developed in the context of nuclear microprobe technologies.[52, 53] Sensitivities of ~30 keV are within reach on the basis of the available state-of-the-art detection technologies. Nonetheless, this requires the development of custom signal amplification electronics with ~fF feedback capacitance and careful reduction of environmental and thermal noise—to be achieved, e.g., through the design of shielded irradiation chambers operating at cryogenic temperatures. Note that this detection approach cannot be employed for target implantation in photonic structures, for these cannot intrinsically be fitted with sensing electrodes. In this case, ion counting could be performed through the detection of secondary electrons. This scheme—yet to be realized for single-ion delivery in diamond devices—would prevent utilizing collimators for nanoscale positioning, as they would absorb the secondary radiation emitted from the target. An alternative path consists in fabricating the photonic structures themselves registered to a set of emitters pre-implanted at specific positions.[54] Other approaches also consider combining ion implantation with in-situ confocal microscopy and annealing to generate on demand emitters. A similar approach was recently realised with laser writing of colour centres in diamond.[55]

*Formation yield.* The incorporation of different ions in diamond is the first step. This is followed by the formation of the implanted impurities into optically-active emitters—i.e. atom-vacancy complexes. This is achieved by annealing at high temperatures, which promotes the recombination of the ions with implantation-induced lattice vacancies. Typical annealing temperatures are in the 800–1200 °C range for all group-IV elements. To date, yields of ≤25%, 1%, 5%, and 2% have been reported for 15-50 keV implantations of Si, Ge, Sn and Pb ions, respectively. These estimates are statistical, due to the limited amount of experiments exploiting single-ion counting techniques. Promising alternative techniques—including for instance high pressure high temperature treatments—might lead to a significant increase in yield. This was demonstrated for other centres in diamond,[56] as well as for the SnV$^-$ with a reported higher stability of the HPHT-treated centres[32] with respect to those subjected to ordinary annealing.[33] Another promising approach consists in combining femtosecond laser-induced annealing with conventional thermal treatments.[57]

**Spin-optical properties**

Key to demonstrate fundamental phenomena, as well as realizing quantum-based devices is the ability to address and manipulate the spin state of these diamond colour centres. This aspect is still at its infancy as is the development, in parallel, of protocols to detect, manipulate and convert the charge state of the M-V defects—for most of these centres exist in multiple charge states. This is particularly important for reducing or ideally eliminating effects such as spectral diffusion and ionisation, which limit the practical employment of these colour centres in quantum applications relying on photon indistinguishability.

**Symmetry and structure.** Group-IV diamond defects display the split-vacancy configuration in which the element M (Si, Ge, Sn, Pb) lies in between two adjacent diamond vacancies, i.e. two missing nearest-neighbours carbon atoms.[13] The centre has nominally $D_{3d}$ symmetry with the diamond <111> axis as the principal, three-fold rotation axis ($C_3$)—though lower symmetries (e.g. $C_2$, $D_2$ point groups) can occur due to lattice strain and deformation. In the negatively-charged state of the defects—which gives the characteristic emission lines in Figure 1b—there are eleven valence electrons involved: six from the dangling bonds of the adjacent carbons atoms, four from the group IV element and one from elsewhere in the lattice. The resulting M-V$^-$ defects have ground ($^2E_g$) and excited ($^2E_u$) states that both have $E$ symmetry and double orbital degeneracy. The orbital degeneracy is lifted by spin-orbit coupling and dynamic Jahn-Teller interaction, leading to a pair of split ground and excited states, each with double spin-degeneracy ($S = 1/2$, Figure 1c).[14,58,59] These eigenstates have corresponding optical and phononic transitions which couple only to the orbital degree of freedom and are spin conserving, allowing for all-optical coherent control schemes of the defect's spin state. Notably, these spin states possess a near-unity spin purity, enabling spin-tagged resonance fluorescence measurements[18] and making these colour centres desirable spin-photon quantum interface candidates for quantum information networks. Further, owing to the $D_{3d}$ symmetry, the application of a magnetic field of given magnitude and direction determines the admixture of the resultant spin eigenstate and can influence the relative relaxation time as the ground and excited state manifolds experience different strengths of spin-orbit interaction. In the SiV$^-$, for instance, tuning the orientation of the magnetic field with respect to the $C_3$ symmetry axis results in the variation of the spin relaxation time from tens of nanoseconds to a few milliseconds—the longest relaxation time being for the field aligned along $C_3$).

**Spin Control.** For many quantum-based applications, the spin coherence of the colour centres is a crucial figure of merit, for it dictates the feasibility as well as the maximum number of fundamental operations executable on the candidate qubit. Ideally, the ratio between the time of a single gate operation and decoherence time of the (spin) state should be $\sim 10^{-3}$–$10^{-6}$, as imposed by error correction limits. Here we focus on the well-studied SiV$^-$ as a representative case; similar considerations though can be drawn for the other diamond M-V centres and are briefly discussed at the end of this section.

For the SiV⁻ the coherence time is relatively short and is thus one of the main challenges towards its use as a robust qubit. The SiV⁻ possesses two possible paths to encode information as a qubit—either via the two orbital branches of the ground state ($\Delta_{GS}$ ~47 GHz)[60] or via the spin $S = 1/2$ sublevels of the orbital states, which have characteristic optical signatures and whose degeneracy is lifted by Zeeman splitting. Compared to the rather long spin relaxation time ($T_1^{spin}$ ~ms), the lifetime of the orbital states of the SiV⁻ centre is limited ($T_1^{Orbital}$ ~tens of ns) due to acoustic phonon scattering between the two ground orbital branches (|12⟩ and |34⟩ in Fig. 1c).[21, 22, 61, 62] The dephasing time involving Zeeman-split spin sublevels ($T_2^*$ ~tens of ns) is also short and of the same order of $T_1^{Orbital}$. It follows the same temperature dependence of $T_1^{Orbital}$, indicating that spin dephasing is also dominated by phonon-mediated transitions between the orbital branches.[22] For completeness, longer values for the spin dephasing time of the SiV⁻, $T_2'$ ~hundred ns, have been reported in Ramsey interferometry experiments measuring the free induction decay time of the spin.[22]
Being such a critical constraint, different approaches have been proposed to increase the coherence time of the SiV⁻, as described below

***Cool-down.*** One option is to cool down the SiV⁻ centre to millikelvin temperatures,[28] which is currently possible with commercial ³He/⁴He dilution refrigerators. At these temperatures, the spin coherence time can reach tens of ms if the magnetic field is well-aligned,[63] as shown in Figure 2a–c. This allows for optical excitations of the SiV⁻ centre for as many as ~$10^5$ cycles before a spin-flip event can occur, thus making high-fidelity single-shot spin readout a concrete possibility, even with low photon collection efficiencies (~$10^{-4}$). At millikelvin temperatures the strong suppression of phonon-mediated dephasing allows for the observation of prominent Rabi oscillations between the two spin ground states. Free induction decay (FID) measurements have shown that the spin coherence time of the SiV⁻ centre can increase from ~tens or hundreds of ns for natural samples[64] to ~10 μs for isotopically-pure diamond (0.001% ¹³C). Extension of the spin coherence time is possible by employing dynamical decoupling protocols to rephase the spin via a series of π-pulses. Notably, in these schemes the fact that the spin coherence time does not saturate with the number of π-pulses indicates that the coherence time can be further extended by reducing the uncertainty on the pulses, for instance by using a decoupling sequence with two-axis control.[65]

***Strain-engineering.*** A second method to enhance the coherence time of the SiV⁻ centre is to manipulate the phonon scattering strength via strain engineering,[66] Figure 2d. Here a tuneable static strain is applied on a SiV⁻ centre that is located in a monolithic single-crystal diamond cantilever with metal electrodes patterned above and below. Voltage applied across the electrodes deflects the cantilever downwards generating a controllable deformation. The strain mixes the orbitals of the ground and excited state manifolds, and alters the splitting ($\Delta_{GS}$ and $\Delta_{ES}$) in each manifold. Theoretical simulations indicate that, at 4.2 K, increasing the orbital splitting $\Delta_{GS}$ would cause the phonon absorption rate of the SiV⁻ centre to first raise slightly (due to the higher phonon density of states), before experiencing a monotonic drop (due to the competing exponential decrease in thermal occupation of the phononic states themselves). Overall, a reduction in phonon absorption causes the suppression of the phonon-mediated dephasing process in the system, effectively increasing the coherence time (Figure 2d).

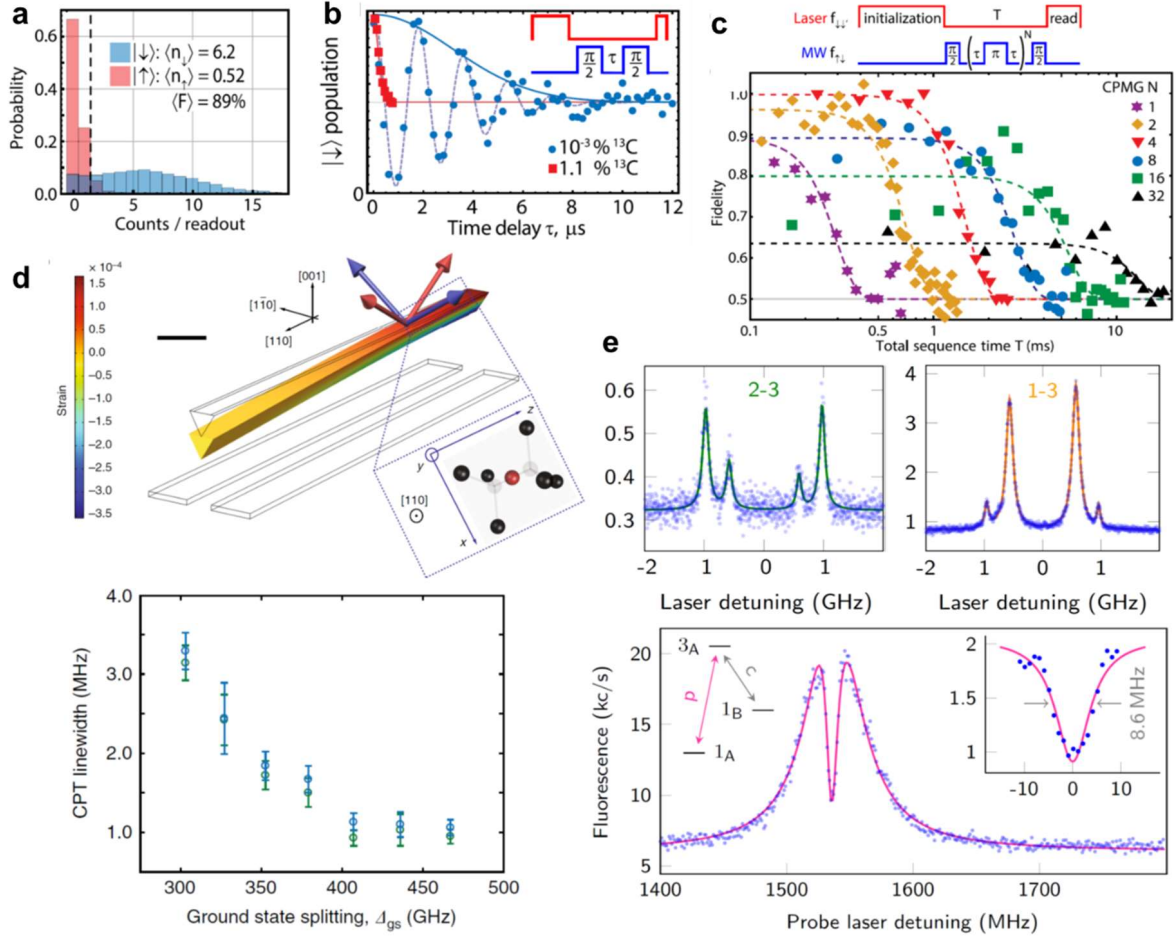

**Figure 2.** Coherent spin control in the SiV⁻ centre. **a)** Spin control of the silicon-vacancy centre at millikelvin temperatures. Single-shot spin readout with magnetic field $B$ = 2.7 kG: a 20-ms-long laser pulse pumping the transition A1 (see Fig. 1c) is used to read out the state after a 250-ms-long initialization pump of the A1 (red) or B2 transition (blue). **b)** Ramsey interference measurement of $T_2^*$ for two samples: ¹³C purified sample (blue, 0.001% ¹³C) and unprocessed natural sample (red, 1.1% ¹³C). The microwave field is detuned by ~550 kHz from the Zeeman splitting between $|1\rangle$ and $|2\rangle$. The duration of the initialization and readout period are 15 ms and 2 ms for the 0.001% ¹³C sample, and 2 ms and 1.5 ms for the 1.1% ¹³C sample. **c)** Carr-Purcell-Meiboom-Gill (CPMG) pulse sequence with $N$ = 1, 2, 4, 8, 16 and 32 pulses in a sample with low ¹³C concentrations and with an aligned magnetic field $B \approx$ 1.6 kG at 100 mK. Durations of the initialization and readout laser pulses are 100 ms and 15 ms, respectively. Dashed lines are fits to $\exp[-(T/T_2)^4]$. **d)** Suppression of spin dephasing via strain engineering. The simulated bending is at an applied voltage of 200 V between top and bottom electrodes. The component of the strain tensor in the direction of the long axis of the cantilever is indicated by the colour scale. Scale bar corresponds to 2 μm. The linewidth of coherent population trapping (CPT) dips as a function of ground-state orbital splitting $\Delta_{GS}$ is shown, indicating an increase in spin coherence for higher levels of strain. **e)** Demonstration of coherent manipulation trapping with GeV⁻ centres. The top panels represent coherent excitation from the lower (1) and higher (2) energy ground state manifold to the excited state (3). A dip with full width at half maximum of (8.6 ± 0.5) MHz is visible, which corresponds to a coherence lifetime of (19 ± 1) ns.

***Charge-state control.*** An indirect alternative is to consider the neutrally-charged state of the M-V defects. For instance, the SiV⁰ displays strong (~90%), nearly transform-limited (~450 MHz) emission in the ZPL at 946 nm. It is a $S$ = 1 spin system which is less subject than its negatively-charged counterpart to decoherence caused by phonon interaction and electric field fluctuations.[67, 68] Spin polarization of the system has been observed,[69] and explained through state-selective intersystem crossing between singlet and triplet states, similarly to what occurs in NV⁻ centres. Additionally, time-resolved electron spin resonance measurements on ensembles of SiV⁰ centres have shown long coherence times of ~1 ms (with Hahn-echo schemes), as well as exceptionally long spin-lattice relaxation times ~40 s, at temperatures <20 K. The dephasing mechanism seems to be dominated by slowly-varying noise from nearby $^{13}$C nuclear spins, which means it can be reduced by improving the quality and isotopic purity of the material.

The main drawbacks of the SiV⁰ are the relatively low fluorescence at low temperature[70] and the technological challenge of controlling the doping of diamond (e.g. controlling the concentration of N and B) to favour the formation of SiV⁰ over that of SiV⁻ centres.

Beyond the SiV⁻ centre, spin manipulation using microwave fields has been demonstrated in GeV⁻ centres.[42] Nevertheless, for the GeV⁻ the ground-state splitting is too small to benefit from suppressing the electron-phonon interaction. This leads to a relatively small spin dephasing time (limited by phonon-mediated orbital relaxation) of ~20 ns—as measured through coherent population trapping experiments at cryogenic temperatures (~2.2 K). The results are shown in Figure 2e. For this measurement, a (100)-oriented diamond was used with magnetic field applied in the plane of the diamond. The larger-than-usual splitting of the ground state is due to the residual strain.[42] The GeV⁻ spin relaxation time was measured to be ~25 μs under precise alignment (54 degrees) between the magnetic field and the GeV⁻ axis. The other M-V centres in the family possess larger intrinsic orbital splitting which suggest they might be less prone to dephasing. In general, these experimental demonstrations show that the family of diamond M-V centres constitute promising systems for developing spin-photon interfaces and quantum networks communicating via photons.

**Quantum Photonics with Group-IV Emitters**

Coherent light-matter interfaces are one of the key components of many proposed quantum technologies. They are for instance essential for advanced quantum networks and modular quantum computing architectures[71] that exploit photon interference to entangle distant long-lived spin-based quantum memories.[72] In this framework, diamond group-IV emitters are a suitable candidate hardware. Their strong emission into the spin-correlated ZPL, symmetry-protected optical transitions, and low inhomogeneous spectral distribution set them above many alternative systems. Further, nanofabrication in the semiconductor industry is well-established, making diamond M-V emitters a concrete route towards realizing integrated nanophotonic devices based on light-matter interactions with single quanta. In this section we review the use of group-IV emitters in some of the main quantum optical technologies.

**Optical properties.** One essential property of a coherent light-matter interface is a transform-limited optical transition. This is because in the opposite case, i.e. when spectral broadening is present, the emitted photons are distinguishable and the fidelity of key linear-optical quantum operations (e.g. projective Bell-state measurements) is limited. A second desirable property is high quantum efficiency: emission of spectrally-coherent, spin-correlated photons into a specific mode with high probability after optical excitation. Finally, the inhomogeneous (emitter-to-emitter) variation in optical transition frequency should be small such that spectrally-identical emitters can be produced scalably. Group-IV defects in diamond have advantages over other solid-state emitters for these crucial aspects. Unlike NV⁻ centres, which have poor optical linewidths when produced through ion implantation,[73] SiV⁻ centres have been shown to have highly coherent optical transitions when incorporated in this fashion. A similarly high photon coherence has since been demonstrated for GeV⁻ and SnV⁻, showing that the inversion symmetry of the diamond M-V defects make them robust in this respect. As mentioned before, diamond group-IV emitters generally have higher Debye-Waller factors than NV⁻ centres, resulting in higher emission efficiency into the ZPL and can have very small inhomogeneous spectral distribution. As a result, in 2014 the experimental demonstration of Hong-Ou-Mandel interference between two SiV⁻ centres was demonstrated.[27] The experiment set a milestone for diamond-based quantum optics and proved that group-IV emitters are serious contenders for advanced solid-state quantum realizations. The key properties of group-IV emitters in this context are summarized in Figure 3 below.

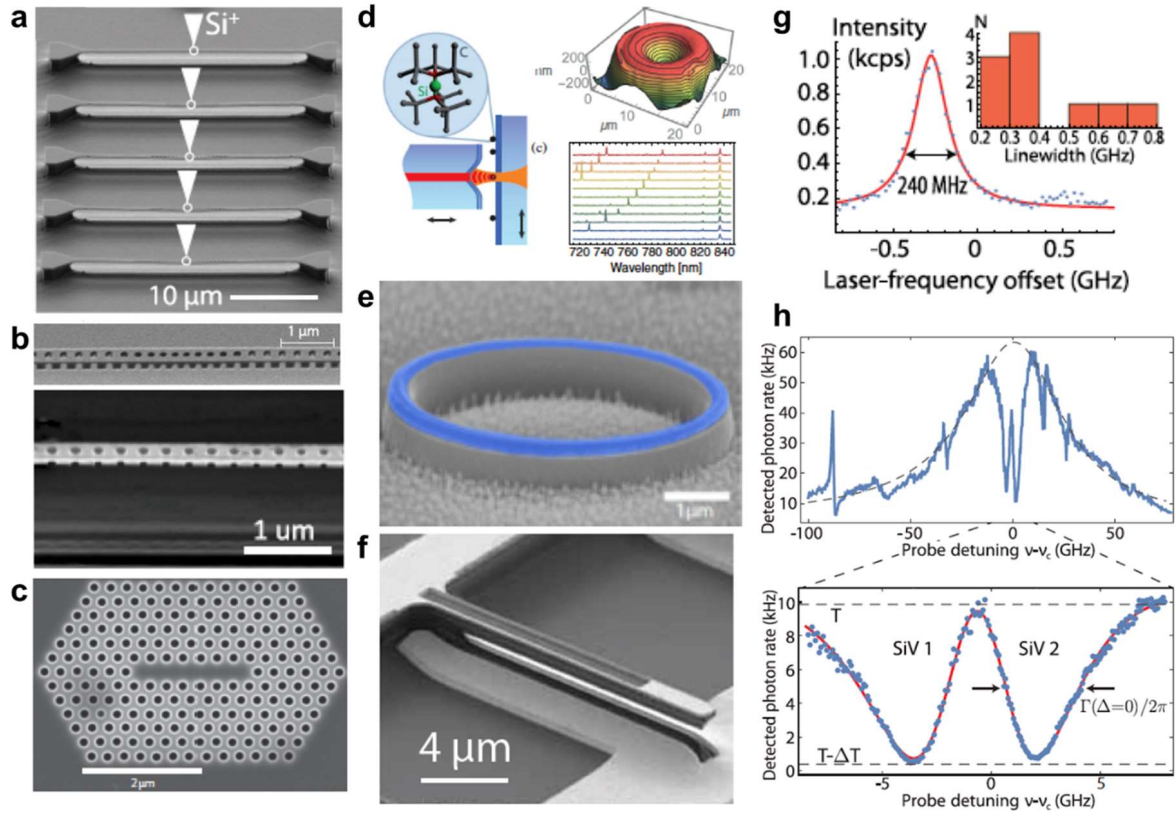

**Figure 3.** Nanophotonic integration of group-IV emitters. Photonic crystals containing group-IV diamond colour centres and including: **a)** 1D triangular nanobeam cavities,[49] **b)** quasi-isotropic etched nanobeams, and **c)** FIB-patterned diamond membranes.[74] **d)** Fibre-microcavity integrated SiV⁻ centres, **e)** GeV⁻ centres coupled to micro-rings, and **f)** group-IV strain tuning devices based on electrostatically-actuated cantilevers. **g, h)** Optical properties of SiV⁻ centres in 1D photonic structures. Th linewidth distribution shows near-lifetime limited SiV⁻ centres in optical structures. Cavity reflectivity measurements indicating a high cooperativity $C > 20$.

**Device Integration.** To further increase the quality of the light-matter interface, the emitters can be integrated into photonic devices.[75, 76] By structuring the electromagnetic environment surrounding the emitter, light-matter interaction can be increased via the cavity-QED cooperativity $C = 4g^2/k\gamma$, where $g$ is the emitter-photon coupling rate, $k$ the loss rate of the optical mode, and $\gamma$ the emitter dephasing rate. The cooperativity is related to the Purcell factor $F_P = C/QE$, with (1-QE) quantifying the emission into all other modes, which increases the emission rate and reduces the emitter lifetime, also reducing the effect of spectral broadening, and to the $\beta$-factor, which quantifies the emission efficiency into a single mode. It also defines optical transmission and reflectivity, key measures of quantum operation fidelities in networking architectures—e.g. the Duan-Kimble scheme. Diamond nanophotonic cavities with high quality factors, $Q > 10,000$, and with low mode volume, $V \sim (\lambda/2n)^3$ have been demonstrated in several realizations.[6] However, the high optical dephasing rate of NV⁻ centres—for which the cavities were designed—near etched surfaces, combined with the high emission of the NV⁻ into the phonon sideband, limited achievable cooperativity to $C < 1$ in these systems.

Nanophotonic integration of group-IV emitters has since overcome these issues. The first demonstration of coupling of group-IV emitters to photonic crystal cavities in bulk diamond dates back to 2012.[74, 77] The SiV⁻ centres were integrated in 1D and 2D FIB-etched structures limited by a relatively low Q of 700 or to a microdisk cavity. In the following years, improved fabrication techniques based on angular[78] and quasi-isotropic[79, 80] oxygen etching led to increased photonic crystal cavity quality factors, resulting in significant advances towards strong-light matter interaction (Figure 3a-c). Other photonic configurations have also been explored, including fibre-based microcavities,[81, 82] where Purcell factors of 9.2 have been shown, and diamond microdisk resonators[83] (Figure 3d, e). At the same time, the use of group-IV emitters has reduced spectral broadening, with near-lifetime limited linewidths reported for SiV⁻ centres in single-mode diamond waveguides[26] (Figure 3g). This combination of improved nanofabrication and reduced spectral width enabled the recent demonstration of high-cooperativity ($C \sim 20$) light-matter interfaces using SiV⁻ centres coupled to 1D photonic crystals,[84] such that high-fidelity gates between single spins and traveling photons[85] are now possible (Figure 3h). Finally, the inhomogeneous spectral distribution of emitters has been overcome through the use of strain tuning, where electromechanical cantilevers are employed to bring multiple group-IV centres into spectral alignment[66, 86] (Figure 3f). With the addition of long spin coherence and high-fidelity control described above, nanophotonic-integrated group-IV emitters have all the properties required for coherent light-matter interface applications.

These exceptional properties have been employed for several advances in quantum optics. For instance, SiV⁻ and GeV⁻ centres in nanophotonic waveguides have been used in demonstrations of superradiant emission, indicating entanglement between spatially separate emitters in a single optical mode (Figure 4a).[49, 87] This set group-IV diamond emitters as candidate fundamental building blocks for quantum networks relying on distant entanglement of quantum nodes. Further work extended this concept to two emitters mutually coupled to a cavity mode.[84] With the increased light-matter coupling afforded by the cavity, coherent interactions can be directly observed via an avoided crossing in the optical spectra as the emitters are magnetically tuned onto resonance with each other (Figure 4b). Strong, controllable coupling as demonstrated in this work can enable deterministic two-qubit gates, an essential element of a quantum repeater node. A third advance is the production of spectrally-tuneable coherent single photons via cavity-assisted Raman processes[88] (Figure 4c). This scheme has the potential for producing indistinguishable single photons in a scalable manner, which is the core requirement for linear-optical quantum computing and network applications. The robust optical coherence and lack of spectral diffusion in group-IV emitters integrated with nanophotonics, including lifetime-limited SiV⁻ and SnV⁻, as well as spectrally narrow GeV⁻, are some of the key elements that enabled these breakthrough demonstrations of light-matter interaction. Further, the demonstrated coherent spin control with nanophotonic-integrated group-IV systems, including ancilla nuclear spin 'data' qubits,[89] sets the concrete possibility for the realization of a fully-functional multi-qubit quantum network node.

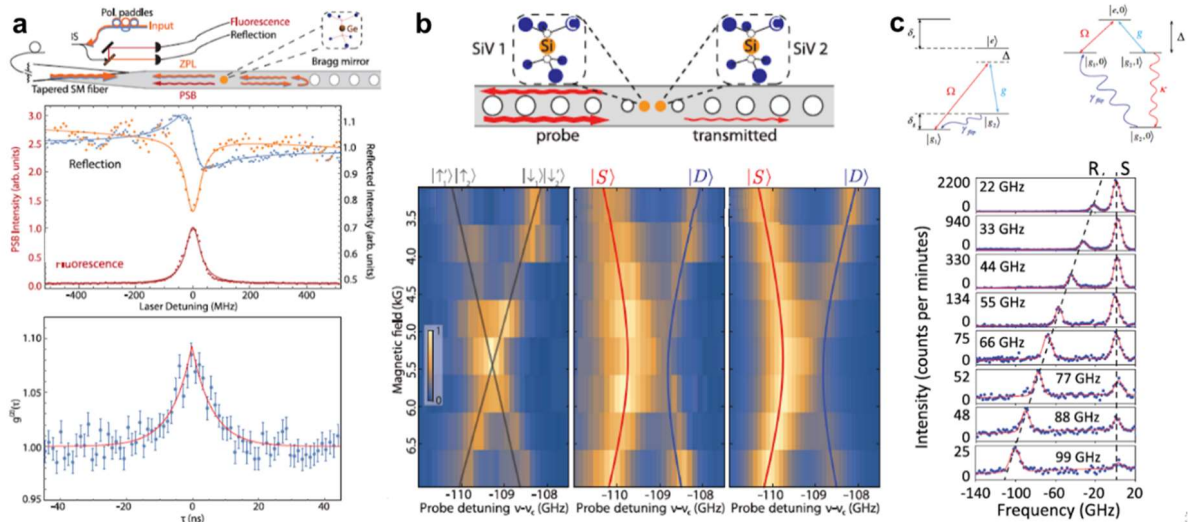

**Figure 4.** Quantum optics with group-IV colour centres. **a)** Superradiant emission of GeV⁻ centres coupled to a single-mode waveguide. The bunching signature with $g^2 > 1$ (lower left) indicates cooperative emission between two separated GeV⁻ centres. **b)** Coherent coupling between two SiV⁻ centres in a photonic crystal cavity. The characteristic avoided crossing (mid panel) indicates a coupling greater than the optical linewidth. **c)** Widely-tuneable Raman emission from a SiV⁻ centre coupled to a photonic crystal cavity.

**Quantum Plasmonics**

Colour centres in diamond have always been one of the most prominent systems for quantum plasmonic experiments owing to their robustness and small size.[90] For instance, NV⁻ centres in nanodiamond have led to the demonstration of wave-particle duality of surface plasmons,[91] and have been successfully coupled to a variety of resonators and waveguides. Remarkably, in a recent experiment a single nanodiamond was positioned in an ultra-small gap cavity and resulted in a two-order-of-magnitude enhancement in luminescence from a hosted single NV⁻ centre, achieving >10 Mcounts/s at saturation at room temperature[92] (Fig. 5a).

Beyond these first realizations exploiting the well-characterised NV⁻ centre, the emergence of the group-IV emitters in diamond—with their narrowband, fully-polarised emission—brought a new wave of experiments as well as novel plasmonic structure designs. Recently a specific geometry involving gold dimers has been proposed to achieve a three-hundred-fold luminescence enhancement from SiV⁻ colour centres in a diamond nanoparticle.[93] Interestingly, the proposed dimensions and geometry are within technological reach as they require a 10-nm nanodiamond coated with few-nm SiO₂ film and located between two gold structures ~65–120 nm in size, depending on the geometry of the dimer (Fig. 5b). The experimental demonstration of this particular proposal is yet to be achieved; nevertheless, the first steps towards on-chip quantum plasmonics with group-IV emitters have already been taken. Figure 5c shows a GeV⁻ centre in a nanodiamond host coupled to a hybrid plasmonic waveguide.[94] The nanodiamond containing a single GeV⁻ centre is embedded in a hydrogen silsesquioxane (HSQ) waveguide on top of a smooth silver film. The photoluminescence signal from the GeV⁻ shows a modest three-fold enhancement and, more importantly, the same scheme allows for excitation of the GeV⁻ remotely, through the grating coupler. Although the waveguide and cavity lengths in this particular example are small (~ tens of µm), longer lengths are technologically feasible with the use of more suitable dielectric crystals (e.g. TiO₂) and larger single-crystal silver films.

New designs and calculations are also being brought forward for realising indistinguishable photon sources at room temperature based on group-IV colour centres in diamond.[95] The basic idea revolves around employing low-Q cavities that have ultra small volume and can lead to strong light confinement and fluorescence enhancement. This approach has already been realised experimentally to demonstrate ultra-bright emission (yet not of coherent or indistinguishable photons) in other systems, including NV⁻ centres, emitters in 2D materials,[96] and quantum dots,[97] with count rates ~MHz. Current losses in the plasmonic waveguides and devices can be minimized by synthesis and growth of better-quality materials, by careful geometric design and improved coupling to the far field. The latter can be achieved, for instance, via a nanoparticle acting both as a scatterer and, simultaneously, as an antenna. In perspective, if the spontaneous emission is decreased to ultimately reach the dephasing limit, experiments in the quantum regime could be performed at liquid nitrogen and perhaps even at room temperature.[95]

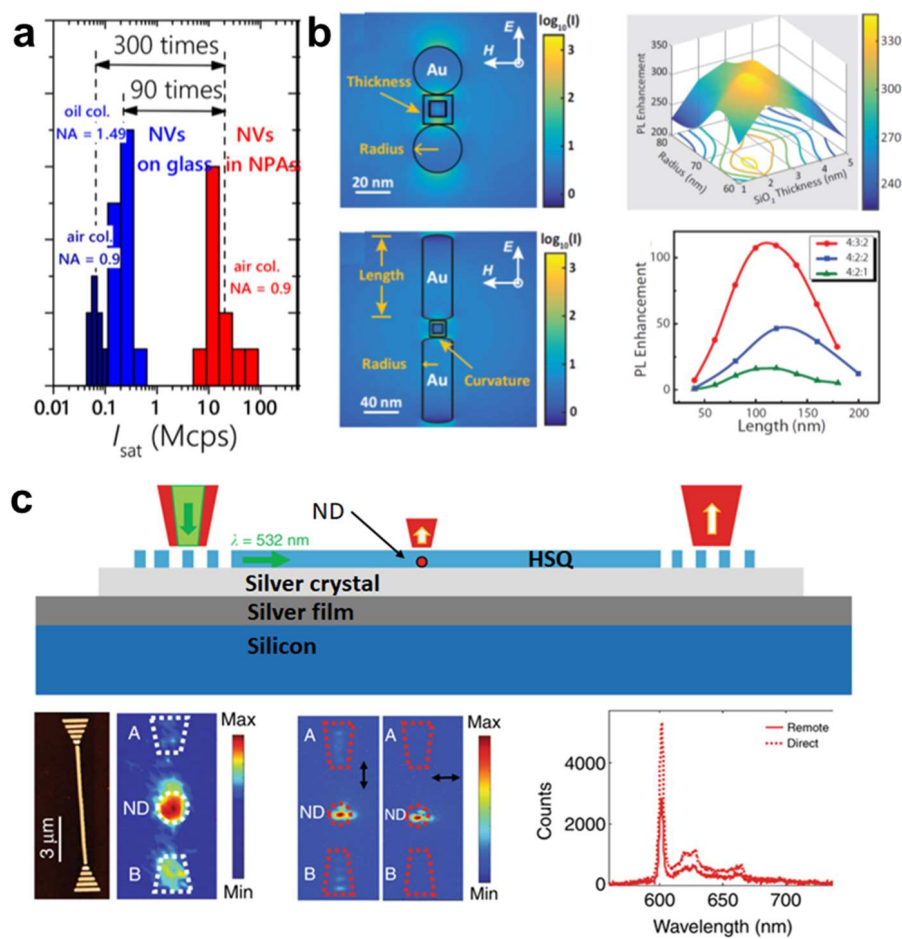

**Figure 5.** Integrated Plasmonics with group-IV defects in diamond. **a)** State-of-the-art enhancement of luminescence from a single NV⁻ centre, resulting in over 1x10⁶ counts/s at saturation. **b)** Modelling of SiV⁻ photoluminescence enhancement in an antenna configuration, showing a potential 300-fold enhancement. **c)** Example of a simple plasmonic waveguide coupled to GeV⁻ centres. Coupling to the waveguide is demonstrated, and remote collection through the waveguide is shown to be better than a direct one.

**Applications, outlook and challenges**

Group-IV diamond emitters have already been successfully employed in a suite of fundamental demonstrations and practical applications. Yet the field is still at its infancy, especially in certain research areas. An in-depth discussion of these is beyond the scope of this perspective, but we deem important highlighting a few notable ones where M-V emitters are showing great promise (Figure 6).

Quantum sensing has been a particularly active area of research for diamond M-V centres. Atom-like in nature, encased in a robust diamond matrix and mere nanometres away from the surface, these colour centres are ideal sensing elements. This feature has been recently explored to measure, for instance, temperatures at the nanoscale via all-optical approaches. The methods rely on monitoring the ZPL amplitude, barycentre or full-width at half-maximum of the emitters as the temperature varies[98-100] or, in more sensitive realisations, on detecting changes in their fluorescence intensity under anti-stokes excitation—which probes the highly temperature-dependent phonons.[101] Given that SiV$^-$ centres can be effectively integrated into atomic force microscopy (AFM) tips,[102] such a method can be a strong contender for all-optical scanning thermometry.

Additional applications that warrant attention in the context of sensing, are labelling and imaging in liquid, particularly with diamond nanoparticles in biological environments. Here, diamond group-IV defects have significant advantages over alternative biomarkers. The diamond particles can be as small as just a few nanometers,[103] they are non-toxic and can be readily functionalized to be target-specific. The hosted M-V emitters have narrowband emission at different wavelengths, which allows for multi-wavelength imaging—nanodiamonds containing different colour centres can be functionalized to target different biological structures and monitored simultaneously through different optical filters. An aspect that is particularly appealing from a practical point of view is in situ-doping during high pressure high temperature growth, which can produce large quantities (~kg) of nanodiamonds. This can be achieved via addition of silicon or germanium-based precursors, that are readily available.

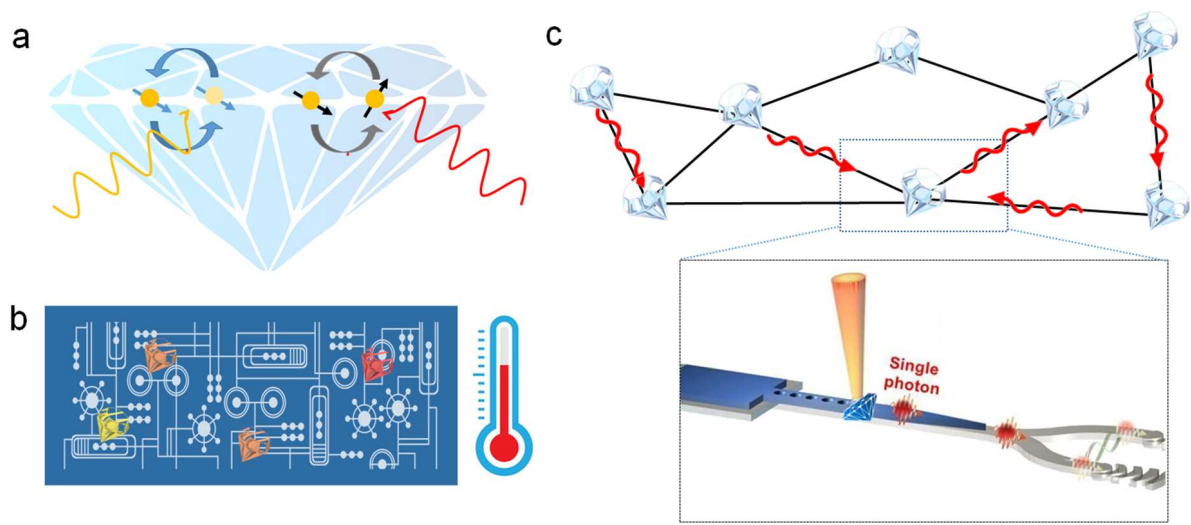

**Figure 6.** Schematic of some of the envisioned applications for group-IV emitters in diamond. **a)** All-optical charge- and spin-state control of group-iV emitters in diamond. **b)** Quantum sensing, e.g. nanothermometry of integrated circuits, using M-V defects in diamond nanoparticle hosts. **c)** Quantum network based on diamond group-IV emitters as spin-photon interfaces. The system exemplified a diamond photonic crystal with a tapered edge coupled to a waveguide made from e.g. aluminium nitride (adapted from Refs[104, 105])

Group-IV emitters in diamond are also showing potential for electroluminescence-based applications where the luminescence of the defect is triggered electrically rather than optically. Unlike NV⁻ centres, SiV⁻ centres can be driven electrically.[106-109] It has been proposed that other group-IV defects could also be excited electrically. This area of research is still at an early stage, but the preliminary results and the forecast predictions are promising. While n-type diamond is a notoriously challenging material to grow, hybrid approaches involving for instance p-type diamond and an external n-type material (e.g. indium tin oxide) can be used as an optical diode.[107] Such control can be exploited to realize scalable, on-chip quantum circuitry—ideally comparable to state-of-the-art GaAs-based quantum dots—as well as to achieve tuning of spin transitions and generation of coherent photons from the colour centres. This is particularly advantageous for the long-term vision of realizing full integration on a single chip, where optical excitation is limited by the diffraction limit.

Another direction where more research is needed is charge-state control. While the negatively-charged M-V⁻ emitters in diamond are well studied, so far only the neutrally-charged SiV⁰ centre has been experimentally observed and characterised. The SiV⁰ is promising, as it offers a combination of long electron spin coherence and desirable optical properties—strong (~90%), nearly transform-limited (~450 MHz) emission in the ZPL in the near infrared at 946 nm. However, engineering the neutrally-charged variant of M-V defects on demand has, so far, been proven challenging as the corresponding negatively-charged defects seem to form more favourably. Current strategies rely on doping the diamond with boron atoms (which can act as acceptors) while limiting the amount of nitrogen atoms (which can act as donors). The precise conditions to achieve a deterministic yield are still under study. There is also an ongoing debate about the feasibility of isolating single SiV⁰ defects in a controllable manner. Whilst expected, currently the neutrally-charged variant of any other diamond group-IV defect has not been observed. Their spectral signature should be offset to the visible range of the spectrum, thus enabling easier and more efficient detection.

Overall group-IV diamond emitters combine desirable optical properties—comparable to quantum dots—and long spin coherence time—similar to NV⁻ centres. We thus also anticipate a bloom of advanced quantum applications requiring spin-photon interfaces. One of the most ambitious realizations consists in the design of a prototype quantum network based on spin-photon and spin-spin entanglement mediated by photon interference. Research on strong coupling between group-IV emitters and cavities, as well as the implementation of spin-photon controlled phase gates mediated by the cavity is currently undergoing, as is the investigation of spin-spin entanglement via direct dipole-dipole coupling and the realization of a qubit platform where nuclear spins are employed as quantum memories.